\documentstyle{cupconf}
\input{psfig}

\def\aa{{\it Astron. Astrophys.} }

\def\aj{{\it A.J.} }
\def\apj{{\it Ap.J.} }
\def\apjl{{\it Ap.J.Lett.} }
\def\apjs{{\it Ap.J.Supp.} }

\def\baas{{\it B.A.A.S}}
\def\mnras{{\it M.N.R.A.S} }

\def\Msun{M_{\odot}}

\def\Lsun{L_{\odot}}

\begin{document}
 
\title[White Dwarfs and the HDF]{White dwarf stars and the {\it Hubble 
Deep Field}}
\author[S.D. Kawaler]{Steven D. Kawaler}
 
\affiliation{Department of Physics and Astronomy, Iowa State University,
Ames, IA 50011 USA}

\maketitle

\section{Introduction}

The value of the {\it Hubble Deep Field} to study of the remote regions of
the observable Universe is difficult to overstate, as the many exciting
results of this workshop are ample testament.  However, an added bonus of the
HDF is that, although it a very narrow angle survey, the depth of the HDF
results in its sampling a significant volume of the halo of our galaxy.  Thus
it is useful for the purposes of detecting (or placing upper limits on the
distribution of) intrinsically faint stars, such as white dwarfs.  In this
capacity, the HDF can say some important things about the possible
constituents of the baryonic dark matter in the halo of the Milky Way, as
well as constrain the physics of white dwarf formation and cooling.

Such an investigation is timely since white dwarfs could provide a
significant fraction of the total mass of the halo of the Milky Way.  The
MACHO observations of microlensing events in the direction of the LMC
suggests that up to 50\% of the dark matter halo of the Milky Way could be
comprised of faint Population II white dwarf stars (Alcock et al.
\cite{macho97}).  Thus, constraints on the population of halo white dwarfs
from the HDF can directly address this possible partial explanation of the
nature of the dark halo of the Milky Way.

In this review, I hope to illustrate how the HDF can be used to constrain the
luminosity function of halo white dwarfs.  I begin with a brief summary of
the observed white dwarf luminosity function (WDLF) of the galactic disk, and
show how the HDF serves as a probe of the WDLF for the halo.  I then review
the theoretical background used in interpreting the WDLF in terms of the
theory of white dwarf evolution and cooling, and the history of star
formation in the galaxy.  We are then in a position to explore the
theoretical WDLF, beginning with the WDLF of the disk.  We then move to a
theoretical examination of the WDLF of the halo population.  The results of
searches for white dwarfs on the HDF can then be examined in terms of the
halo white dwarf population.

\subsection{White dwarf stars}

White dwarf stars represent the final stage of evolution of stars like 
our sun.  They represent the ultimate fate of all stars with masses less 
than about 8 $\Msun$, and are the natural consequence of the finite fuel 
supply of these stars.  Upon exhaustion of their nuclear fuels of 
hydrogen and helium, these stars lack sufficient mass to take advantage 
of the limited return of carbon fusion, and are doomed to gravitational 
collapse.  Upon reaching a radius comparable to that of the Earth, the 
inner cores of these stars support themselves almost entirely by electron 
degeneracy pressure.  Their final transformation is through the gradual 
release of heat that has been stored within them during the prior stages 
of nuclear burning.

There is a simple and compelling reason to care about white dwarf stars.  It
can be shown that {\it 98\% of all stars are or will be white dwarfs!} In
terms of stellar mass, 94\% of all matter in stars is either already
locked-up in white dwarfs, or is in stars that will eventually become white
dwarfs; a mere 6\% of matter is destined to be, or already incorporated in
neutron stars, and an insignificant amount is on the black hole path.

The basic theory of white dwarf stars was stimulated by their identification, 
early in this century, as a class of stars with observable luminosity 
but very small radius.  The solution to this mystery, by Chandrasekhar 
and Fowler, neatly integrated the then--new sciences of quantum 
mechanics and relativity with astrophysics.  Their work was 
essentially a complete description of the inner workings of cool white 
dwarf stars.  While Chandrasekhar's theory is indeed elegant, in the 
later half of this century we have come to recognize that these stars 
represent a rich storehouse of information on the evolution of all 
stars.  This information is available through examination of the luminosity
function of white dwarf stars.

The relatively recent discovery of white dwarfs is evidence that they are
relatively hard to find.  Almost all white dwarfs that we know of are in the
immediate solar neighborhood; 50\% of known white dwarfs lie within 24
parsecs of the Sun.  White dwarf discoveries have come from surveys of stars
with large proper motions, searches for faint blue objects, and examination
of faint members of common proper motion pairs.  All three produce candidate
objects that require spectroscopic follow-up observations for confirmation.
Spectroscopic follow-up has identified a large number of white dwarf stars.
However, the selection effects involved in such surveys are thorny to account
for, and make statistical studies suspect.

The most successful recent surveys for producing new white dwarf 
identifications have been surveys for faint but excessively blue stars.  A 
prime example is the Palomar--Green (PG) survey (Green et al.  1986) for 
blue objects out of the galactic plane.  While the PG survey (and others 
like it) are designed for discovery of quasars, they are almost optimally 
designed to pick out white dwarf stars.  Once again, spectroscopic 
follow-up is required, but such surveys have an advantage in that they 
produce magnitude--limited samples, and therefore are without the peculiar 
selection effects associated with proper--motion identifications.  With the 
success of the the PG survey, others are underway, such as the 
Montreal--Cambridge survey (Demers et al. \cite{demetal86}) and the 
Edinburgh--Cape survey (Stobie et al. \cite{stobetal87}, Kilkenny et al.
\cite{kilketal91}).  

Because of the relatively narrow range in masses of white dwarfs (and their
confinement to a line of constant radius in the H--R diagram) the Greenstein
colors allow a fair determination of the absolute visual magnitude $M_v$.  
In this system, the (U-V) color provides a useful indicator for the temperature
of hot white dwarfs, while the (G-R) color index works best for cooler white
dwarfs.  Conversion of $M_v$ to bolometric magnitude $M_{bol}$ requires model
atmosphere bolometric corrections, such as given in Greenstein (\cite{gree76})
for cool hydrogen--atmosphere white dwarfs.  Of course, when parallax
measurements are available, more accurate estimates of individual values of
$M_v$ and $M_{bol}$ are possible.

\subsection{The observed white dwarf luminosity function and the age 
of the Galactic disk}

Using estimates of $M_v$, one can construct a luminosity function for 
white dwarfs.  The luminosity function is a representation of the 
relative space densities of white dwarfs with given absolute magnitudes, 
usually plotted in terms of number of stars per cubic parsec per unit 
bolometric magnitude, versus luminosity.  This luminosity function for 
low luminosities by Liebert et al. (\cite{liebetal88}) used data on 
stars in the Luyten (\cite{luyt79}) survey.  An example of the 
observed luminosity function is shown in Figure 1, with data from
(\cite{liebetal88}).  The luminosity function increases steadily with 
decreasing luminosity.  As could be expected, the cooler a white dwarf 
is, the more slowly it cools and fades, and so the number increases 
with lower temperature and luminosity.

\begin{figure}
\psfig{file=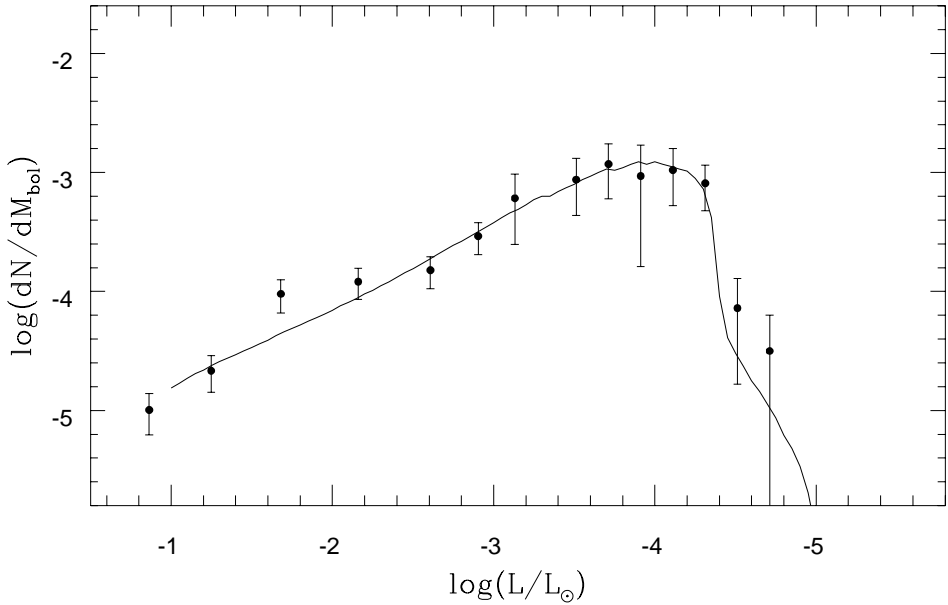,height=9.0cm}
\caption[]{The luminosity function of DA white dwarfs.  Data points are from
Fleming et al. (\cite{flemetal86}) and Liebert et al. (\cite{liebetal88}). 
The line is a representative theoretical luminosity function using the pure
carbon white dwarf models of Winget et al. (\cite{wingetal87}) and a disk 
age of 8 Gyr.}
\end{figure}

However, there is a steep turn--down in the luminosity function below
$10^{-4.4}\Lsun$ that requires explanation.  This turn--down is the result of
the finite time that white dwarfs in the solar neighborhood, and by extension
in our galaxy, have had to cool.  The time it takes for white dwarfs to fade
to below this luminosity must be longer than the age of the galaxy itself.
Thus the luminosity of this cut--off is a direct measure of the age of our
galaxy.  For Figure 1, I show a sample theoretical luminosity function
calculated with simple models of pure carbon white dwarfs published by Winget
et al.  (\cite{wingetal87}).   Matt Wood's results (\cite{wood92}) using the
best available white dwarf models and observed luminosity functions, indicate
that the age of the galactic disk in the vicinity of the Sun is $9.3\pm 1.5$
Gyr.  From the shape of the luminosity function, it is apparent that the star
formation rate has been roughly constant over the history of the disk.  Had
there been bursts of star formation at some times, these bursts would have
produced bumps along the observed luminosity function that are not seen in
the data (Wood \cite{wood92}, Iben \& Laughlan \cite{ibelau89}).

\subsection{Other ways of finding white dwarfs: the HDF and MACHO searches}

The techniques described above are adequate for studying relatively bright
white dwarfs that lie reasonably close to the Sun.  As such, they are
targeted for white dwarfs from the disk of the galaxy.  Halo white dwarfs, on
the other hand, could escape detection in these surveys by several means.  If
they are not close to the Sun, such halo white dwarfs could easily be fainter
than the magnitude limits of these surveys.  With the high velocities
expected for halo stars, halo white dwarfs brighter than the limiting
magnitude of current surveys could have enormous proper motions; thus they
would have escaped detection by having proper motions larger than the upper
limit of proper motion studies.

The HDF represents one approach to finding halo white dwarfs: 
extremely deep surveys over a small range of the sky that can find 
very faint stars.  The discussion below (from Kawaler \cite{kawa96}) 
shows that the HDF samples a reasonable volume of the halo.  Another 
novel approach is to search for microlensing events, where lensing of 
distant stars occurs through the action of an intervening halo white 
dwarf.

\subsubsection{The halo as probed by the HDF}

For a small survey area such as the HDF, the volume $V$ of space sampled 
out to a distance $d$ (in pc) given an area of $A$ square arc minutes is 
simply
\begin{equation}
 V = 2.82\times 10^{-8} Ad^3 \; \; {\rm pc}^3 \; .
\end{equation}
The HDF covered an area of approximately 4 square arc minutes; therefore
since it looked out of the plane, taking $d=500$~pc suggests that it samples
a disk volume of only 14~pc$^3$.  With such a small volume sample, it is not
surprising that no disk white dwarfs are seen in the HDF; the approximate
space density of white dwarf stars is $3\times 10^{-3}$ per cubic parsec
(Liebert et al.  1988).

For a putative population of white dwarfs in the halo, the HDF has sampled a
much larger volume.  Assuming a limiting magnitude of $m_l$, and a white
dwarf with an absolute magnitude $M_f$ (in the same band as the limiting
magnitude), then the effective volume contained within can be written in
terms of the limiting magnitude of the survey and the absolute magnitude of
the faintest white dwarf:
\begin{equation}
\log(V)= -4.550 + \log(A) + 0.6(m_l-M_f)
\end{equation}
where $V$ is in pc$^3$ and $A$ is in square arc minutes.  Using some
representative numbers for the HDF, $m_l \approx 27.5$ and $M_f \approx 16.5$
yields a search volume of 450 ${\rm pc}^3$.

Now, assume the density distribution of the halo is spherically symmetric
about the center of the galaxy, and follows a standard ``softened potential''
distribution such as described in Binney \& Tremaine (\cite{bintre87}, p.
601, with $\gamma=2$).  With this form for the halo mass distribution, and
parameter for the halo as in Bahcall \& Soneira (\cite{bahson80}), the mass
of the halo sampled by the HDF field (in solar masses) can be be computed
(for the details, see Kawaler (\cite{kawa96}).  The halo mass sampled by the
HDF ranges from 1.05$\Msun$ to 7.8$\Msun$ for a reasonable range of $M_f$ 
and $m_l$.  Thus the HDF in principle samples several solar masses of halo 
material.

\subsubsection{MACHO searches}

The MACHO collaboration has published an analysis of 8 microlensing events in
the direction of the LMC, which they attribute to halo objects in the Milky
Way (Alcock et al. \cite{macho97}).  The duration of these events suggests
that the lensing objects have masses between 0.1 and 1.0 $\Msun$.  The
lensing objects must be of extremely low luminosity; therefore ordinary dwarf
stars are ruled out.  Lensing at this rate allows the MACHO collaboration to
estimate that up to 50\% of the dark matter halo of the Milky Way could
therefore be comprised of white dwarf stars.  Considering the large mass of
the galactic halo, this implies that halo white dwarfs must be extremely
abundant.  Direct observational constraints on the halo white dwarf
luminosity function from observations (Liebert et al. \cite{liebetal88}) are
not necessarily in conflict with the MACHO result; however the luminosity
function of white dwarfs in the disk must then contain a fair fraction of
halo white dwarfs.

Can the lenses responsible for the MACHO results be white dwarfs, and still
be consisted with the white dwarf space density found by Liebert et al.
(1988)? Given the age of the halo as determined from, for example, globular
cluster studies, the observed downturn in the white dwarf luminosity function
along with the absence of any white dwarfs with $M_{\rm bol} > 16.2$
constrains the star formation history for the generation of halo stars that
might have produced such a halo white dwarf population (Tamanaha et al
\cite{tamaetal90}, Adams \& Laughlin \cite{adalau96}).  Under conventional
assumptions about star formation early in the history of the galaxy, Adams \&
Laughlin (\cite{adalau96}) conclude that the observations of Liebert et al.
(\cite{liebetal88}) already limit the fraction of the dark matter halo that
can be attributed to white dwarfs to less than 25\%.  On the other hand,
Tamanaha et al.  (\cite{tamaetal90}) show that extreme conditions, such as an
enormous burst of star formation early in the history of the galaxy, could
produce a massive number of white dwarfs in the halo that could escape
detection by traditional ground--based studies (see below).

\section{The basics of white dwarf cooling}

To understand the WDLF requires combining the star formation rate as a
function of time and mass, the processes by which main sequence stars become
white dwarfs, and the rate at which white dwarf stars, once formed, fade and
cool.  Therefore, as Don Winget is fond of saying, the history of star
formation in the disk of our galaxy is written in the coolest white dwarf
stars.  In this section, I concentrate on one element of the theoretical
white dwarf luminosity function: the theory of white dwarf cooling.  For more
details see, for example, Kawaler (\cite{saasfee97}) and Van Horn
(\cite{vanh71}) and references in between.  With an understanding of white
dwarf cooling, the observed luminosity function can constrain the other
inputs.  In the next section, we will combine white dwarf cooling with the
remaining inputs needed to model the complete WDLF.

\subsection{Simple ``Mestel'' cooling theory}

In a remarkable 1952 paper, Leon Mestel laid the groundwork for the study 
of white dwarf cooling (Mestel \cite{mest52}).  Following the discovery of 
the peculiar nature of white dwarf stars, the question arose as to what 
their power source might be.  Nuclear burning, identified as the energy 
source in ordinary stars, was an early suspect, but Ledoux \& 
Sauvenier--Goffen (\cite{ledsau50}) showed that if white dwarfs were 
powered by nuclear burning, then they would be vibrationally unstable.  
Since at that time white dwarfs were known to be non--variable, their 
conclusion was that another mechanism was needed.  Mestel (\cite{mest52})
identified the luminosity source as release of stored thermal energy.

The reservoir of thermal energy stored in the core of a white dwarf 
star is slowly depleted by leakage into space.  The 
cooling process is not unlike that undergone by a heated brick placed in a 
cool environment.  Such a brick would cool rapidly when exposed to the cool 
outside air unless surrounded by some sort of insulating 
blanket.  With insulation, the cooling of the brick is slowed as the 
temperature gradient between it and its surroundings is reduced by the poor 
thermal transport properties of the blanket.  In a white dwarf 
star, heat is transported through the degenerate core by the efficient 
process of electron conduction, while in the nondegenerate envelope energy 
is transported by the (much less efficient) form of photon diffusion.  Thus 
in a white dwarf star, the degenerate core corresponds to the hot brick, 
and the nondegenerate outer layers play the role of an insulating blanket.

Mestel (\cite{mest52}) began with the equations of stellar evolution, and
then made several simplifying assumptions for the case of white dwarf stars.
Nuclear energy generation and gravitational contraction are assumed to play
no role.  The specific heat in the electron-degenerate, isothermal
(temperature $T_c$) core is set by the (nondegenerate) ions.
Combining these assumptions, and integrating through the degenerate core, the
luminosity of a white dwarf star can be expressed in terms of the total
stellar mass, core composition, and rate of change of the core temperature.

Atop the degenerate core is the nondegenerate envelope, which contains (by
assumption) only a small fraction of the mass of the star.  If the envelope
is radiative, then the rate of energy transfer is governed by the radiative
opacity of the material.  Adoptin a Kramers opacity law and also assume a zero
boundary condition leads to the so--called radiative zero solution for $T$ as
a function of $P$ (see for example Hansen \& Kawaler \cite{hankaw94}).
Matching this solution to the degenerate core yields an expression relating
the stellar luminosity to $T_c$ and the core composition.

Combining these two expressions for the luminosity and integrating with
respect to time yields the time needed to cool (actually, fade) to a given
luminosity:
\begin{equation}
t_{\rm{cool}}=9.41\times 10^6 \: {\rm yr} \left(\frac{A}{12}\right)^{-1}
              \left(\frac{\mu_e}{2}\right)^{4/3} \mu^{-2/7}
              \left(\frac{M}{\Msun}\right)^{5/7}
              \left(\frac{L}{\Lsun}\right)^{-5/7}
\end{equation}
Some features to note about this remarkable result include the dependence of
$t_{\rm cool}$ on $A$ and $M$.  Cores with larger $A$ (cores composed of
heavier elements) cool faster than ``lighter'' cores.  If we assume
representative values for mass ($0.60\Msun$), atomic weight ($A=14$, a 50/50
mix of carbon and oxygen), $\mu=1.4$, $\mu_e=2$, and a luminosity
corresponding to the faintest white dwarfs ($L=10^{-4.5}\Lsun$), then this
simplified analysis results in a cooling time of $7\times 10^9$ years.

\subsection{Complications at the cool end: crystallization}

This estimate of the cooling time is remarkably close (within 30\%) of the
most modern computation of the cooling age of these coolest white dwarf
stars.  This despite the fact that it leaves out several effects that are now
known to be very important: neutrino cooling, prior evolutionary history,
crystallization effects, etc.  Van Horn (\cite{vanh71}) reviews the Mestel
theory and evaluates the precision of the assumptions; Iben \& Tutukov
(\cite{ibetut84}) discuss the Mestel cooling law in light of their more
detailed evolutionary models.

The Mestel law assumes that the white dwarf interior is an ideal gas, 
in which there are no electromagnetic interactions between the nuclei.  
But this is, after all, material in the core of a white dwarf star, where 
densities are enormous by terrestrial standards.  Above some density 
(and at finite temperature) the ions are indeed sufficiently crowded 
that their electrostatic interaction can affect the equation of state.
As white dwarfs cool, the effects of Coulomb interactions become more
important.  Also, since the interiors of white dwarfs are roughly isothermal,
the Coulomb effects are largest at the center.  More details relevant to
white dwarf interiors are available from numerous sources; see for example
Shapiro \& Teukolsky (\cite{shateu83}) and references therein.

When the density becomes large enough, Coulomb effects overwhelm those of
thermal agitation, long--range forces organize the small-scale structure of
the material, and the gas settles down into a crystal.  For conditions
relevant to white dwarf stars, with central densities of order $10^6 [\rm
g/cm^3]$, oxygen crystallizes at about $3.4\times10^6K$, while carbon
crystallizes at $2.1\times10^6K$.  These central temperatures are reached
when the white dwarf has been cooling for about $10^9$ years.  Thus
crystallization can be an important process in the evolution of white dwarf
stars.  Note that the crystallization condition assumes a one component
plasma, while material within most white dwarfs is probably a mixture of at
least carbon and oxygen.  The actual process of crystallization must be
extremely complex.

\subsubsection{Effects of crystallization on the cooling rate}

As a white dwarf cools, the specific heat for the degenerate interior changes
in ways most familiar to condensed matter physicists.  Here, I only sketchily
summarize the behavior of the specific heat; further details (in language
appropriate for astrophysicists) are clearly described by Shapiro \&
Teukolsky (\cite{shateu83}).  In an ideal perfect gas, such as the ions
within a white dwarf, the specific heat at constant density is independent of
temperature, and proportional only to the mean atomic weight $\mu$.  In the
limit of complete degeneracy (such as electrons in a white dwarf core), the
internal energy is independent of the temperature.  For such a system, the
specific heat is identically zero.  Therefore, as white dwarf material
becomes degenerate, the specific heat of the electrons becomes very small.
Since the ions are nondegenerate, however, the specific heat of the ions
remains unchanged.  Therefore the specific heat of electron--degenerate
material is determined solely by the specific heat of the ions which is a
constant for a given composition.

When crystallization begins, the latent heat release associated with the
growing long--range ordering provides an additional luminosity source.
Lattice vibrations (in three dimensions) ultimately result in a doubling of
the specific heat as the material cools towards crystallization.  With a
higher specific heat, the rate of change of the luminosity decreases with
time as the latent heat of crystallization provides an additional source of
energy.  In average white dwarfs, this occurs at $\log L/\Lsun \approx -3.6$
to $-4.2$.  In this luminosity range, a white dwarf at a given luminosity is
probably older than the Mestel law would indicate.

As the core cools further, the internal energy of the lattice is determined
by quantum mechanical effects.  Upon cooling below the Debye temperature, the
material reaches the Debye cooling phase, with a specific heat that drops in
proportion to the temperature.  The rate of change of luminosity increases
dramatically; a low specific heat means that the star cannot hold in thermal
energy easily.  Thus once a white dwarf cools below the Debye temperature,
its luminosity plummets faster than the Mestel law alone would predict.  In
this Debye cooling phase, a white dwarf at a given luminosity is probably a
bit younger than the Mestel law would suggest.

Figure 2 shows a representative white dwarf cooling curve, kindly provided by
Matt Wood, as a solid line.  The dashed line represents Mestel
cooling. At the low luminosity end, the bump corresponding to the release of
latent heat of crystallization is clearly evident. The lowest luminosity
portion shows the effects of Debye cooling

\begin{figure}
\psfig{file=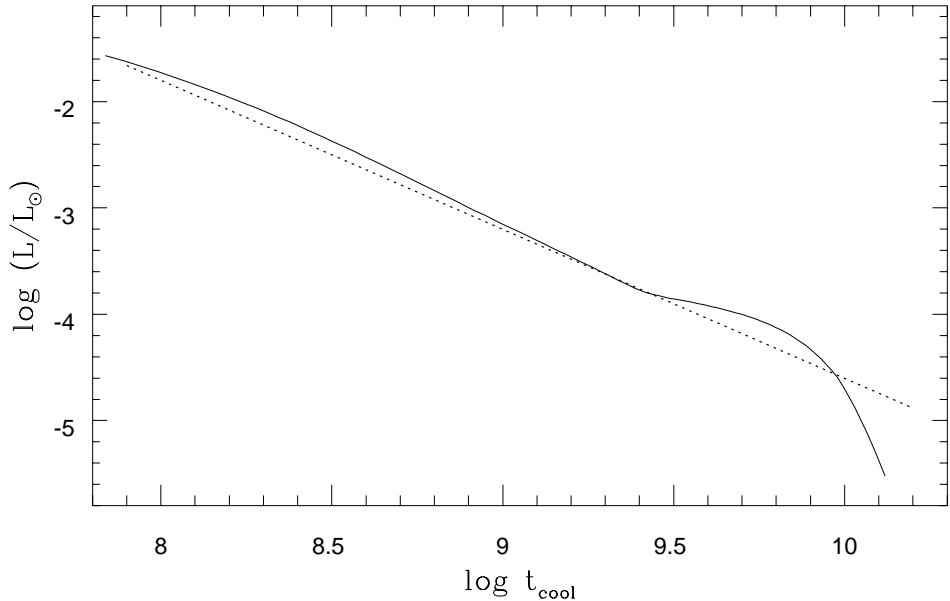,height=9.0cm}
\caption[]{A representative cooling curve for a  0.60$\Msun$ white dwarf model 
(solid line) with a carbon core, and stratified outer layer of helium
($10^{-2}M_{\odot}$ and hydrogen ($10^{-4}M_{\odot}$), kindly provided by Matt
Wood.  This model includes the effects of crystallization, but not
fractionation (since it is pure carbon in the core).  The dashed line
is the comparable cooling curve based on the Mestel law.  The differences
illustrate how the Mestel cooling law is modified by inclusion of
crystallization effects.}
\end{figure}

\subsubsection{Fractionation during crystallization: another energy 
source?}

Along with the latent heat of crystallization, another possible 
source of energy in a cooling white dwarf is rooted in the fact that 
the material undergoing crystallization is of mixed composition.  Most 
white dwarf stars have degenerate cores composed of a mixture of 
carbon and oxygen, along with at least trace abundances of other 
heavier elements.

The temperature at which material may crystallize depends on the atomic
weight and charge of the material.  As indicated earlier, the temperature at
which oxygen can crystallize is approximately 50\% higher than the
temperature at which carbon does so.  Thus considered separately, oxygen will
begin to solidify within the still (degenerate) gaseous carbon environment.
What will happen to the growing oxygen crystalline material?  Will it
condense out forming precipitating grains that sink towards the stellar
center?  Will the material remain mixed but ``slushy'' until it cools below
the crystallization temperature for carbon?

The question of differential crystallization and possible fractionation is a
very important one.  If it occurs in a way that causes element separation,
then the gravitational potential energy released during the settling acts as
an energy source that can slow the cooling of the star as a whole.  If
plotted in Figure 2, the effects would be to stretch out the solid curve to
the right (longer times) at low luminosities.  Currently a very active area
of research, the situation is explored in some detail by Segretain et al.
(\cite{segretal94}) and others (see for example Isern et al. \cite{iseretal97}),
who conclude that such a process can slow the evolution of carbon/oxygen white 
dwarf stars to the lowest observed luminosities by up to two billion years 
or more.

\subsection{Realistic calculations}

In constructing realistic cooling curves, modelers of white dwarf stars try
to include all of the known physics within the model.  The physical
properties enter the cooling curve as the constitutive relations that provide
the coefficients of the equations of stellar structure and evolution.  These
models use realistic conductive and radiative opacities, sophisticated
equations--of--state, including the effects of degeneracy, Coulomb
interactions, and crystallization.

Three representative sequences from the calculations of Wood (as quoted by 
Winget et al.  \cite{wingetal87}) are shown in Figure 3.  This figure 
shows several of the basic scalings of the Mestel law hold for these 
complete models.  First note that at a given luminosity the more massive 
models are older, as expected from the above equation.  All of the tracks 
parallel the Mestel law; that is, they show a general power--law slope of
\begin{figure}
\psfig{file=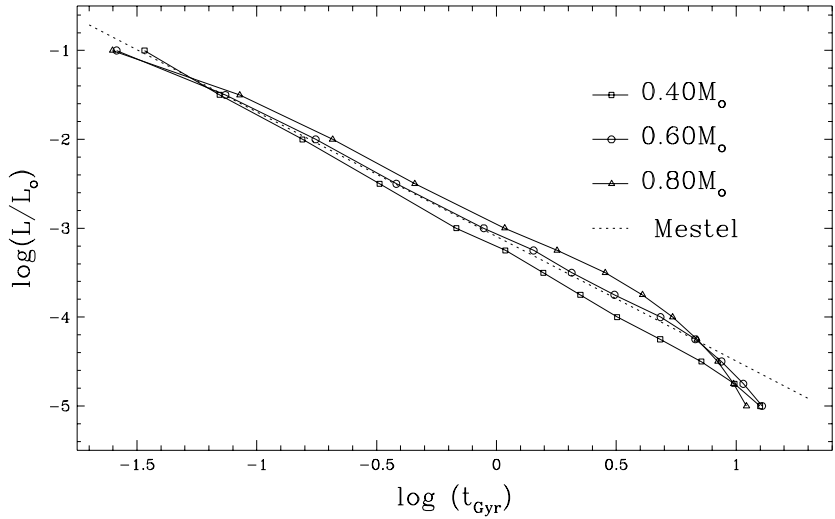,height=8.4cm}
\caption[]{Cooling curves for representative white dwarf evolutionary 
sequences.  Data from Winget et al. \cite{wingetal87}; figure from Hansen 
\& Kawaler \cite{hankaw94}}
\end{figure}
-1.4.  Note that the luminosity of the $0.80 \Msun$ model shown begins to 
drop quickly at a luminosity below $10^{-4}\Lsun$.  This is the effect of 
crystallization; these models have already largely crystallized, and they 
are showing the effects of Debye cooling discussed in a previous section.  
The $0.60 \Msun$ model begins Debye cooling at a lower luminosity and later 
time; this implies that more massive models cool more slowly, but that they 
crystallize earlier.

Before the downturn, the cooling curve flattens compared to the Mestel law.  
This is consistent with the expectations, described above, that as 
crystallization begins, the specific heat rises with the release of latent 
heat of crystallization.  This ``extra'' luminosity source slows the 
evolution.  Once largely crystallized, the models enter the Debye cooling 
stage resulting in the acceleration of the luminosity drop of the white 
dwarf models.  Because of this, even though they cool more slowly initially, 
we expect that the faintest white dwarfs in the galaxy have, on average, 
higher mass than slightly more luminous (and therefore younger still) white 
dwarfs.

The most complete models of cooling white dwarfs, and the white dwarf 
luminosity function, is the work of Wood (\cite{wood92}), who computed 
many sequences of evolving white dwarf models with many different 
combinations of parameters.  Since white dwarf stars at the cool end of 
the luminosity function have largely forgotten their initial conditions, 
Wood used a homologous set of starting models.  Convergence of the 
evolutionary tracks was essentially complete by the time the models 
reached a luminosity of roughly $0.1\Lsun$.

\section{The WD luminosity function of the disk}

\subsection{Constructing theoretical WDLFs}

With theoretical cooling curves, we need a way to see if the 
theoretical inputs to the models are realistic.  The principal 
observational contact of the theory of white dwarf cooling is the 
observed {\it luminosity function} of white dwarf stars: the number of 
white dwarfs per cubic parsec per unit luminosity (or bolometric 
magnitude), usually denoted as $\Phi$.

The cooling rate of white dwarfs is one input into the construction of a
theoretical luminosity function.  As described by Wood (\cite{wood92}), a
theoretical determination of the luminosity function requires evaluation of
the expression
\begin{equation}
\Phi(L) = \int_{M_{\rm low}}^{M_{\rm hi}}
          \: \psi(t) \: \phi(M) \: \frac{dt_{\rm cool}}{d \log (L/\Lsun)}
                         \: dM 
\end{equation}
at a given value for the population's age.  In this expression, $\psi(t)$ 
is the star formation rate at the time of the birth of the white dwarf 
progenitor, and $\phi(M)$ is the initial mass function for stars with 
initial mass $M$.  The mass limits of the integration cover the mass 
range of main sequence stars that produce white dwarfs.  The upper 
mass limit is simply the largest main--sequence mass that produces 
white dwarfs (about 8-10 $M_{\odot}$).  The lower mass limit is 
approximately the main--sequence turn--off age for the population age; 
simply put, stars lower than this mass have not yet produced white 
dwarfs.  One may obtain this mass limit by estimating the main 
sequence lifetime with a relation such as
\begin{equation}
\log t_{ms} = 9.921 - 3.6648(\log M) + 1.9697(\log M)^{2} - 
0.9369(\log M)^{3}
\end{equation}
from Iben \& Laughlin (\cite{ibelau89}.
The initial mass function (IMF) of stars in the galaxy, $\phi_s(M) dM$ 
is the number of stars formed per year per cubic parsec within an 
interval of masses between $M$ and $M+dM$.  Salpeter (\cite{salp55}) 
found that
\begin{equation}
\phi\, dM = 2\times 10^{-12} M^{-2.35}\,dM \;\; {\rm stars/yr/pc^3},
\end{equation}
which is the famous ``Salpeter mass function''.  For the disk, one begins by
assuming a constant star formation rate $\psi_o$ with a value adjusted to
match the observed white dwarf space density.

The remaining quantity is the inverse of the rate of change of the 
white dwarf luminosity; it is a time scale for luminosity change that 
we'll denote as $\tau_{\rm cool}$.  This number is provided by the cooling 
theory for white dwarfs given the white dwarf luminosity and mass.  In 
the form above, however, $\tau_{\rm cool}$ needs to be specified at a 
given white dwarf luminosity and progenitor mass $M$.  To determine 
$\tau_{\rm cool}$, one needs to know the mass of the white dwarf that a 
star with a main--sequence mass $M$ produces, the time it takes to 
cool to luminosity $L$, and the fading rate at that luminosity.  The 
white dwarf mass corresponding to an initial mass $M$ comes from 
empirical studies of the $M_i - M_f$ relation, which has been explored 
extensively for the past two decades.  A useful parametric form is 
presented by Wood (\cite{wood92}) as
\begin{equation}
M_{f}=0.49 \exp(0.095 M_{i}) \; .
\end{equation}
Given this mass for the white dwarf, and the luminosity $L$, white 
dwarf cooling models can be consulted to yield the cooling age 
$t_{cool}$, its derivative, and therefore $\tau_{\rm cool}$.

Iben \& Laughlin (\cite{ibelau89}) show how the luminosity function appears 
for several simplified relations for the star formation rate, etc.; this 
paper is an excellent starting point for those who wish to work with the 
white dwarf luminosity function for various purposes.  Figure 4
shows an example of a simplified luminosity function and the 
effect of the population age on it.  Quite simply, the older the population
\begin{figure}
\begin{center}
\psfig{file=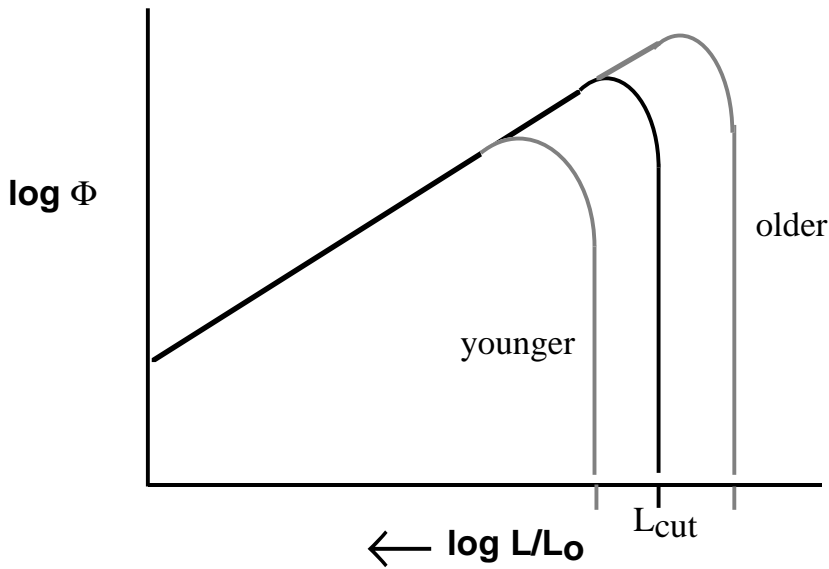,height=7.5cm}
\end{center}
\caption[]{Schematic luminosity functions for populations of three 
different ages, from Kawaler (\cite{saasfee97}).  This illustrates the 
case of a constant star formation rate with time, a Mestel cooling law 
($t_{\rm cool} \propto L^{-5/7}$), and three disk ages.}
\end{figure}
the lower the luminosity of the cutoff.  The stars populating the lowest 
luminosity bin are the oldest white dwarfs in the sample; with increasing 
age, they reach lower luminosity and pile up because of the lengthening 
cooling time scale.

When crystallization begins, the cooling rate temporarily slows, and the 
luminosity function will rise above the Mestel curve.  Debye cooling will 
cause a drop in $\Phi$ with the associated accelerated cooling.  Of course, 
realistic calculations of the luminosity function are needed for detailed 
comparison with the observed luminosity function.

Figure 5 shows a sample of realistic luminosity functions constructed using
realistic white dwarf models computed (and kindly provided) by Matt Wood
(described in Oswalt et al. \cite{oswaetal96}).  Note that the general slope
of this cooling curve is nearly $-5/7$, but with a depression at high
luminosities and a bump near the final cutoff.  Also shown in Figure 5 is the
observed luminosity function of Liebert et al. (\cite{liebetal88}).
Agreement between the theoretical and observed luminosity functions above the
drop--off is very good; those portions that disagree may result from changes
in the star formation rate and/or the IMF over time.  Such possibilities
provide an intriguing use of the white dwarf luminosity function to explore
the history of star formation in our galaxy (or, in the near future, in other
stellar populations such as globular clusters).

Figure 5 clearly shows the effect of increasing age on the white dwarf 
luminosity function; older populations have fainter white dwarf stars.  The 
low--luminosity cutoff decreases in luminosity with increasing population 
age; the figure shows luminosity functions for ages from 7 Gyr to 12 Gyr.  
Inspection of this figure reveals that the observed luminosity function is 
consistent with an age for the disk of the Milky Way of between 8 and 11 
Gyr, with a best value of about 9 Gyr.
\begin{figure}
\psfig{file=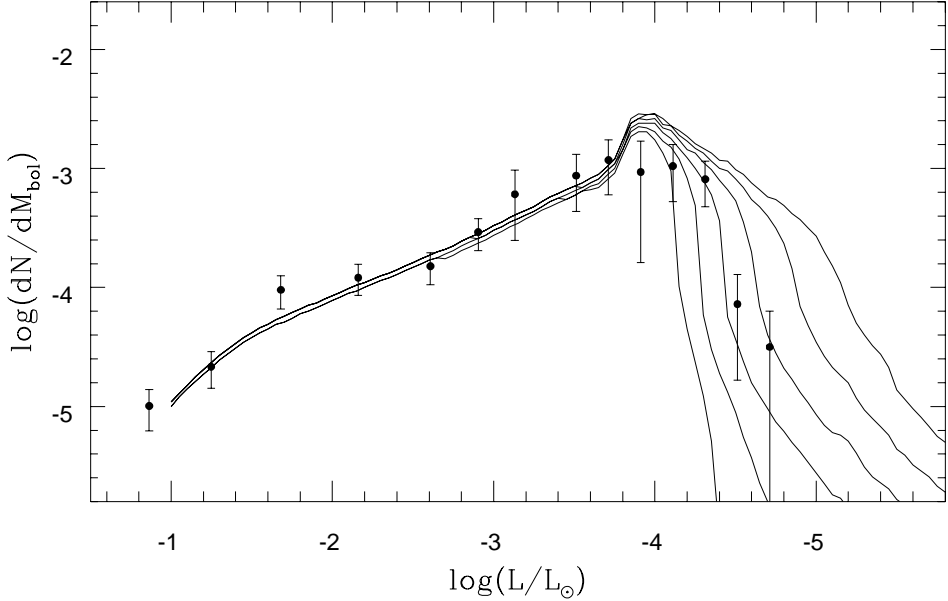,height=9.0cm}
\caption[]{Theoretical white dwarf luminosity functions (constructed with DA
white dwarf evolutionary sequences computed by Matt Wood) for several input
disk ages. The drop--off in luminosity corresponds to disk ages from 7 Gyr
(at the highest luminosity) to 12 Gyr.  Also shown is the observed white
dwarf luminosity function.}
\end{figure}

\subsection{Sensitivity to the input physics}

The derived age of the galactic disk depends on the input physics used in 
the computation of the white dwarf cooling rates.  Given an observed, or 
otherwise fixed, luminosity function for comparison, the uncertainties in 
the derived ages follow from uncertainties in various properties of the 
white dwarf models.  These properties, the sensitivity of the age to them, 
and the real range of possible ages, are summarized in Table 1.  This 
table uses information from Winget \& Van Horn (\cite{winvan87}) and Wood 
(\cite{wood92}).  It lists how the time for a white dwarf with a mass 
of $0.60\Msun$ to drop to a luminosity of $10^{-4.4}\Lsun$ changes with 
various changes in the input physics.
\begin{table}
\begin{center}
\begin{tabular}{l@{\hspace{0.5cm}}c@{\hspace{0.5cm}}l}
Input physics (``$X$'')  & $\frac{d t_{\rm disk}}{d \log X}$ & real range\\
 & & \\
$m_{\rm helium}$        & -0.7 Gyr     & $\leq 1.4^*$ Gyr \\
$A_{\rm core}$ (C? O?)  &  -16 Gyr     & $\approx 2^*$ Gyr \\
conductive opacity       & +7.6 Gyr     & $\leq 0.3$ Gyr \\
radiative opacity        & +1.4 Gyr     & $\leq 0.06$ Gyr \\
$Z_{\rm env}$            & +1.4 Gyr     & $\approx 0.2$ Gyr\\
fractionation            & --           & $< 2 $ Gyr \\
other stuff              & ?            & $< 1 $ Gyr ? \\
 & & \\
\multicolumn{3}{r}{$^*$ measurable with pulsation observations}
\end{tabular}
\end{center}
\caption{Dependence on $t_{\rm disk}$ on input physics}
\end{table}

For example, if the mass of the surface helium layer is increased by one 
decade, the cooling time for a $0.60\Msun$ white dwarf will decrease by 0.7 
Gyr.  In this tabulation, the values of $m_{\rm helium}$, $A_{\rm core}$, 
and $Z_{\rm env}$ are uncertain because of unknowns in the prior evolution 
of the white dwarf stars --- they depend on how the star became a white 
dwarf star, on the $^{12}C(\alpha,\gamma)^{16}O$ nuclear reaction cross 
section, and on the trace metal content in white dwarf envelopes.  The 
``real range'' represents the current uncertainty in the cooling time given 
current uncertainties in the listed parameters.  The dominant uncertainties 
arise from the thickness of the surface helium layer and on the core 
composition.  The combination of these uncertainties yields a current 
best--estimate of the age of the galaxy of about $9.3\pm 1.5$ Gyr.

Fortunately, through observations of the pulsating white dwarf stars, there
is a way to measure these quantities that is independent of approaches that
use the observed luminosity function (see Kawaler \cite{saasfee97} for
details).  These observed pulsations allow us to measure the depth of
subsurface transition zones and, therefore, the thickness of the surface
helium layer in the pulsating DO and DB white dwarfs.  Other pulsators place
constraints on the core composition through the rate of period change.

\section{The white dwarf luminosity function of the halo}

As described earlier, the white dwarfs that are likely to appear on 
the HDF are halo white dwarfs.  This section discusses the 
issues that need to be addressed to examine such a halo population.

\subsection{Differences from the disk component}

Nearly all of the work in the field so far has been on the white dwarf 
luminosity function of the Galactic disk, simply because that is where 
the observed white dwarfs live.  When considering the white dwarf 
component of the galactic halo, however, many of the inputs into 
constructing theoretical WDLFs must be changed, based on what we know 
(or think we know) about stellar properties and star formation during 
the initial collapse of our galaxy.

\subsubsection{Star formation rate: a burst}

Whereas star formation has continued in the disk from the early history of
the galaxy to the present, the stars of the halo formed over a very brief
period very early on.  Therefore, in modeling the WDLF for halo stars, a
nearly universal assumption is that halo stars formed in a single burst at
some time in the past (Tamanaha et al.  \cite{tamaetal90}).  The time of this
burst is the age of the oldest stellar component of the Milky Way.

With this assumption, the star formation rate $\psi(t_{h})$ becomes 
simply a constant, with a value equal to the total number of stars (of 
all masses) produced in the burst.  Associated with this is the time 
$t_{h}$ of the burst; adjusting $t_{h}$ allows examination of how the 
WDLF depends on the halo age.

This simplifies the numerical calculation of the WDLF for the halo; if we
consider the star formation rate such a $\delta$ function, then 
\begin{equation}
-\frac{dN}{dM_{\rm b}} = 
\frac{\log(10)}{2.5}\; \psi(t_h)\; \phi(M)\; \tau_{\rm cool}
         \left[\frac{dt_{\rm ms}}{dM_{\rm i}}+\frac{dt_{\rm cool}}{dM_{\rm wd}}
\frac{dM_{\rm wd}}{dM_{\rm i}}\right]^{-1}
\end{equation}

\subsubsection{The initial mass function}

Several lines of evidence point to an IMF for the halo 
that was quite different than the IMF of the disk.  The observed 
scarcity of low--mass halo stars requires an IMF for halo stars (i.e.  
Bahcall et al.  \cite{bahcetal94}) that does not continue to rise at 
low mass (as does the Salpeter IMF for the disk).  At the high-mass 
end, the disk IMF applied to the halo would have produced a higher 
metallicity (through supernovae) in halo stars than is observed (Ryu 
et al.  \cite{ryuetal90}).  Thus the IMF for the halo is a function 
that is peaked at an intermediate mass.

Additional theoretical evidence leads to an IMF that is parameterized 
conveniently, following Adams \& Laughlin (\cite{adalau96}), as
\begin{equation}
\ln \left[\frac{dN}{dM} (\ln M)\right]
= A-\frac{1}{2 \sigma^{2}}\left[\ln \left(\frac{M}{m_{c}}\right)\right]^{2}
\end{equation}
where the parameter $A$ is a normalization parameter, $\sigma$ characterizes
the width of the distribution, and $m_{c}$ is related to the mass at the peak
of the distribution.  Adams \& Laughlin (\cite{adalau96}) choose base values
for these parameters are $m_{c}=2.3$ and $\sigma=0.44$, though allowable
values range from $m_{c}$ of 2.0 to 4.0 and $\sigma$ ranging from 0.10 to
0.40 for a halo population.  They also show that for a disk population, the
Population I IMF has values for these parameters closest to $\sigma=1.57$ and
$m_{c}=0.15$.

\subsubsection{The initial--final mass relation}

The initial--final mass relation was derived empirically using Population I
stars, typically in moderate--aged open clusters.  Theoretical initial--final
mass relations have reproduced the observed relation by imposing mass loss on
the Asymptotic Giant Branch using a variety of uncertain mass--loss
prescriptions.  Because of the dependency of the initial--final mass loss
relation on such mass loss, it is likely that the relationship will be
different for the metal--poor progenitors of the halo white dwarfs.

Such low--metallicity stars will have envelope opacities that may be
significantly smaller than their Pop I counterparts.  The most likely
mechanism driving mass loss on the AGB is radial pulsation (such as in Mira
variables; see Bowen and Willson {\cite {bowwil91}).  If so, mass loss rates
on the AGB for Pop II stars may have been smaller... leading to white dwarfs
growing to much larger masses inside of them than Pop I stars.  This would
steepen the initial--final mass relation, and bring the lower--mass limit for
Type II supernovae well below the 8 to 10 $M_{\odot}$ that it is for
Population I.

\subsection{Models of the halo WDLF}

We now have all the ingredients we need to explore the shape of the WDLF for
halo white dwarf stars.  For more details about some of the results described
in this section, see  Tamanaha et al. (\cite{tamaetal90}), Adams \& Laughlin
(\cite{adalau96}), Chabrier et al. (\cite{chabetal96}), and Graff et al
(\cite{grafetal98}).  Here we assume a total number density of halo white
dwarfs of $4\times 10^{-3} {\rm pc}^{-3}$, which corresponds to a mass
density of approximately 25\% of the dark halo.

Considering first the halo WDLF in isolation, the influence of the IMF on the
WDLF is shown in Figure 6.  For this figure, I used Wood's DA white dwarf 
models to compute the halo WDLF using the equation shown above.  All three
curves correspond to a halo age of 12 Gyr, but with different values for the
parameters governing the IMF.   The heavy line shows the WDLF corresponding
to the reference values of $m_c=2.3$ and $\sigma=0.44$.  Narrowing the IMF by 
reducing $\sigma$ results in a narrower WDLF (compare the heavy and light 
lines in Figure 6), while increasing the value of $m_c$ shifts the WDLF to
lower luminosities.  For reference, the dotted line shows the WDLF
using the IMF for the disk.
\begin{figure}
\psfig{file=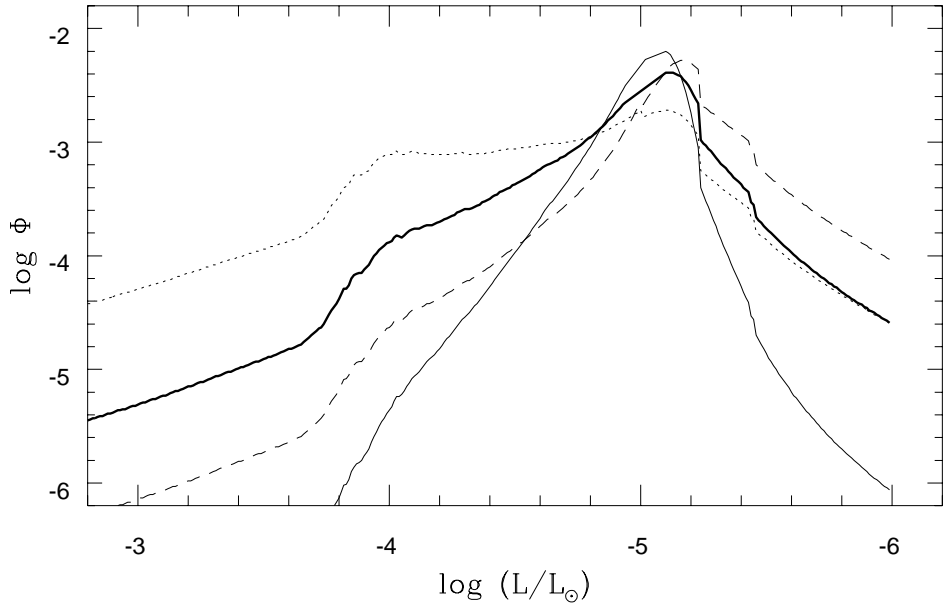,height=9.0cm}
\caption[]{Theoretical WDLFs for halo models at 12 Gyr.
The heavy solid line corresponds to an IMF with parameters $m_c=2.3$ and
$\sigma=0.44$.  A narrower IMF ($\sigma=0.24$) is shown as a thin line, and
an IMF centered at higher mass is shown by a dashed line ($m_c=3.3$). The
dotted line shows the halo WDLF assuming a disk IMF, for comparison.}
\end{figure}

The effect of the halo age on the WDLF of halo white dwarfs is similar to
that of increasing $m_c$.  Older halos produce WDLFs that reach lower
luminosities with broader peaks.  This is clearly a result of the fact that
the white dwarf population has had more time to cool to lower luminosities.
The entire distribution moves towards lower luminosity.  In Figure 7, the
evolution of the WDLF for a halo white dwarf population is clearly evident.
\begin{figure}
\psfig{file=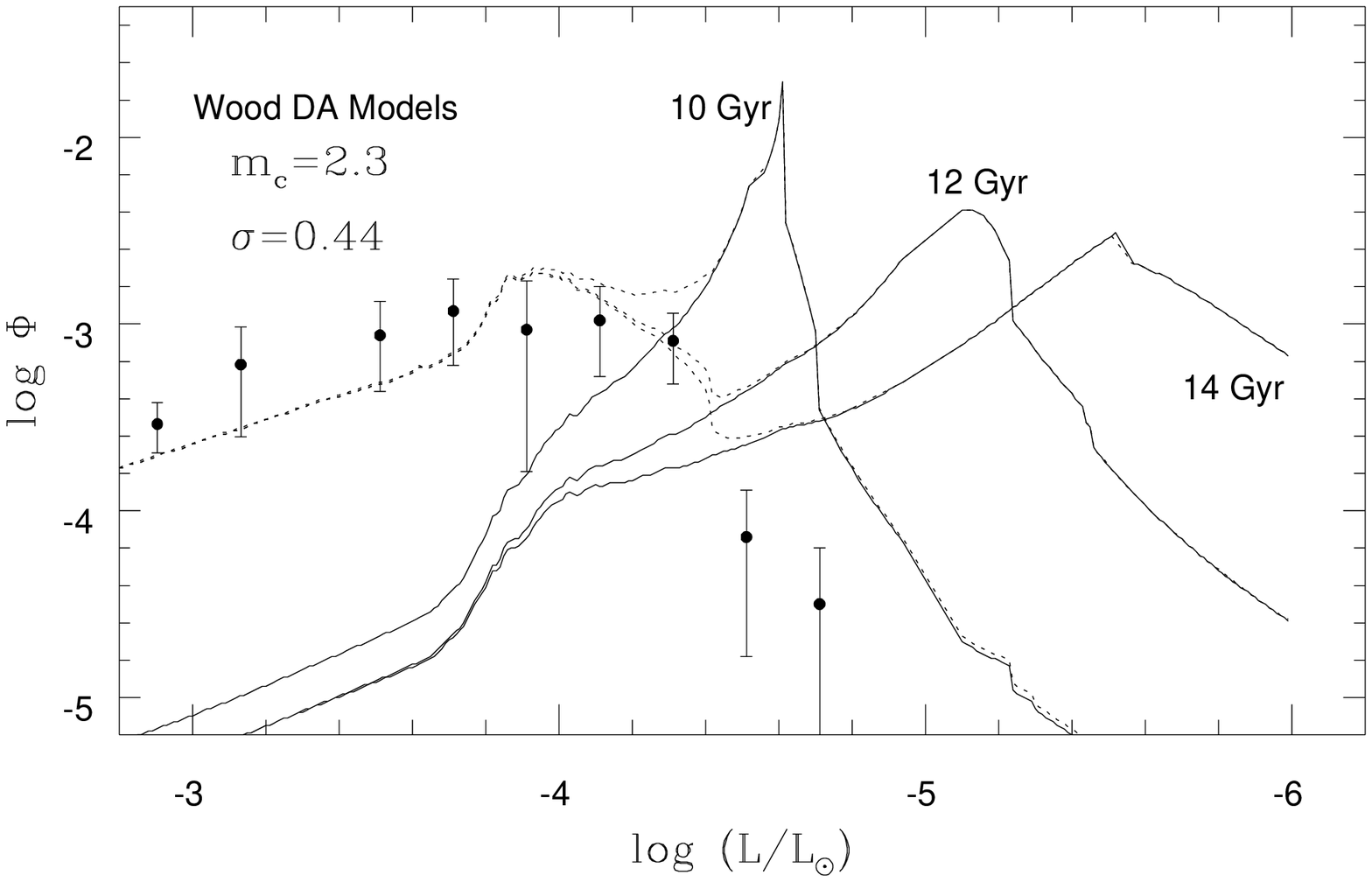,height=9.0cm}
\caption[]{Theoretical WDLFs for halo models with fixed IMF parameters, and
the same total space density, at ages of 10, 12, and 14 Gyr.  The dotted
lines are the complete WDLF including disk stars. The Liebert, et al.
(\cite{liebetal88}) observed disk WDLF is also shown.}
\end{figure}
Under these circumstances, the older the halo, the greater the chance that
halo white dwarfs may have escaped detection by current surveys.  In
addition, if the IMF for the halo is skewed towards higher mass stars at the
expense of low mass stars (i.e. a larger $m_c$) then the halo portion of the
composite WDLF will be forced to lower luminosity.

More complete discussions of the form of the halo WDLF can be found in the
papers cited at the beginning of this subsection; Figures 6 and 7 should
serve to illustrate the main points about the dependence of the halo WDLF on
age and the IMF.  For example, Adams \& Laughlin (\cite{adalau96}) explore a
range of values of the parameters of the IMF (using pure carbon white
dwarf models) -- and the consequences for other observables including the
residual gas left over after the mass loss that halo stars must have
undergone in the production of halo white dwarfs.  Graff et al.
(\cite{grafetal98}) consider the effects of fractionation during
crystallization on the halo WDLF, as well as the many selection effects
associated with surveys, and the issue of the bolometric correction for such
cool objects.

\section{Observational constraints on halo white dwarfs}

The MACHO results have already been discussed, and are provocative indeed.
There is a firm upper limit on the number of halo white dwarfs: the total
mass of halo white dwarfs must not exceed the mass of the galactic halo!  At
or near the position of the Sun, then, the halo white dwarf density must be
less than about $1.7\times 10^{-2} {\rm pc}^{-3}$.  This is a very large
number that exceeds (by nearly an order of magnitude) the density of white
dwarfs currently known.  Interestingly, if halo white dwarfs are to account
for a significant fraction of the halo mass, then the number density cannot
be significantly lower than this.  Thus if the MACHO interpretation of their
data is correct, such white dwarfs must have escaped detection by current
surveys.

\subsection{Presence (or absence) of halo white dwarfs in existing 
white dwarf surveys}

How do studies of the white dwarf luminosity function that were restricted to
the solar neighborhood constrain the {\it halo} WDLF?  This issue has been
addressed by Adams \& Laughlin (\cite{adalau96}) who show how the luminosity
function of white dwarfs in the solar neighborhood as reported by Liebert et
al. (1988) limits the number density of halo white dwarfs.  The last data
point in the disk WDLF is an upper limit of approximately $6\times
10^{-5}$pc$^{-3}M_{\rm bol}^{-1}$ at $\log L = -4.7$.  For luminosity
functions that derive from standard models of cooling white dwarfs, Adams \&
Laughlin (1996) show that their luminosity function must quickly rise to
several $\times 10^{-4}$pc$^{-3}M_{\rm bol}^{-1}$ and remain high down to
very low luminosities if the age of this population is not excessively larger
than the age of the oldest globular clusters.  

This is apparent from examination of Figure 7, which compares the WDLF to a
``standard'' computation of the WDLF including a halo population.  At $\log L
= -4.7$, the theoretical WDLF lies above the observed upper limit for all
halo ages, using the most likely parameters of the IMF.  Recall that these
theoretical WDLF calculations assume that the total mass of halo white dwarfs
is 25\% of the total halo mass.  To be consistent with the last WDLF point
from Liebert et al. (\cite{liebetal88}), the contribution of white dwarfs to
the mass of the halo must be much less than 25\%.

The fraction of the halo mass contributed by white dwarfs can be greater if
the width of the IMF is much narrower than $\sigma=0.44$, as shown in Figure
6.  If $\sigma=0.24$ then a halo age of 14 Gyr can have a white dwarf
contribution of 25\% and still be consistent with the lowest luminosity WDLF
point.  Thus the IMF width must be extremely narrow for the halo
white dwarfs to contribute significantly to the mass of the halo and to have
largely escaped detection in traditional searches for white dwarfs.  This one
of the essential points made by Adams \& Laughlin (\cite{adalau96}) and Graff
et al. (\cite{grafetal98}).

With this constraint alone, the interpretation of lensing events in the MACHO
project as white dwarfs requires a halo age of 14 Gyr or more along with a
rapid rise in the luminosity function at lower luminosities.  Can this
prediction be addressed by the HDF?   Read on...

\subsection{White dwarfs (or lack thereof) on the HDF}

If, as suggested by the MACHO results, the halo dark matter of the Milky Way
is up to 50\% (by mass) halo white dwarfs, then up to half of the halo mass
sampled in the HDF can be white dwarf stars.  The mass sampled is dependent
on the minimum absolute magnitudes of these halo white dwarfs and the
magnitude limit of the images.  Kawaler (\cite{kawa96}) computes the expected
number of white dwarfs on the HDF at any magnitude with minimal assumptions
about the WDLF and the magnitude distribution of halo white dwarfs.
Following Adams \& Laughlin (\cite{adalau96}), this section will describe
more precisely the way that the WDLF of the halo is affected at different
luminosities by the result of searches for white dwarfs on the HDF.

Despite heroic efforts at finding them, there are few obvious stars in the HDF
images (apart from a few ``bright'' 20th magnitude stars); the task of
discriminating between stellar and nonstellar objects at very faint
magnitudes requires extreme care (see the review by John Bahcall in this
volume).  For example, Flynn et al. (\cite{flynetal96}) report that no
white dwarfs exist down to $V=26.3$ for objects with $2.5 > V-I > 1.8$.
They do find bluer stellar objects ($V-I$ between 0 and 1.8) that are
consistent in number as well as color with low--mass main sequence stars.
Similar results have been reported by Mendez et al. (\cite{mendetal96}).
Thus it appears that there are no white dwarfs on the HDF down to $V<28$ or
so.  As discussed by Flynn et al. (\cite{flynetal96}) and by Bahcall (these
proceedings), confident discrimination of true stellar objects from compact
faint galaxies is not possible at higher magnitudes on the HDF.

While disappointing, this upper limit still important when trying
to constrain the white dwarf population of the halo.  We can use it to
further constrain the shape of the WDLF at very low luminosities.  Here we
use equation (1.2) from the Introduction; this equation shows the volume
sampled by the HDF given these figures, and allows calculation of an upper
limit to the space density of white dwarf stars at each luminosity bin.  For
simplicity, we consider as an upper limits to the space density
$1/V$, with $V$ determined with equation 1.2, and zero bolometric 
correction (but see Graff et al.  \cite{grafetal98}).  This then gives the 
upper limit for white dwarfs based on the HDF null result as
\begin{equation}
\log \Phi_{\rm up} = -6.80 - 0.6 m_{V, {\rm lim}} 
              -1.5 \log\left(\frac{L}{L_{\odot}}\right)\;.
\end{equation}
For each magnitude fainter that searches for white dwarfs on the HDF can go,
the WDLF upper limit drops by 0.6 dex at a given luminosity.

Figure 8 shows the HDF limit on the halo contribution (as a dotted line,
assuming a limiting magnitude of $V=28$) to the WDLF along with several 
models of the WDLF and the data for disk white dwarfs.   In Figure 8, as
earlier, the models are for a halo white dwarf population with a total mass
equal to 25\% of the mass of the dark halo.  Each panel represents model WDLFs
with the same disk age (9 Gyr) and halo ages of 12 Gyr and 14 Gyr, with the
14 Gyr luminosity function lying to the right in all three panels.  As can be
surmised by examining Figure 7, a halo age of 10 Gyr is clearly inconsistent
with the data in all cases.
\begin{figure}
\begin{center}
\psfig{file=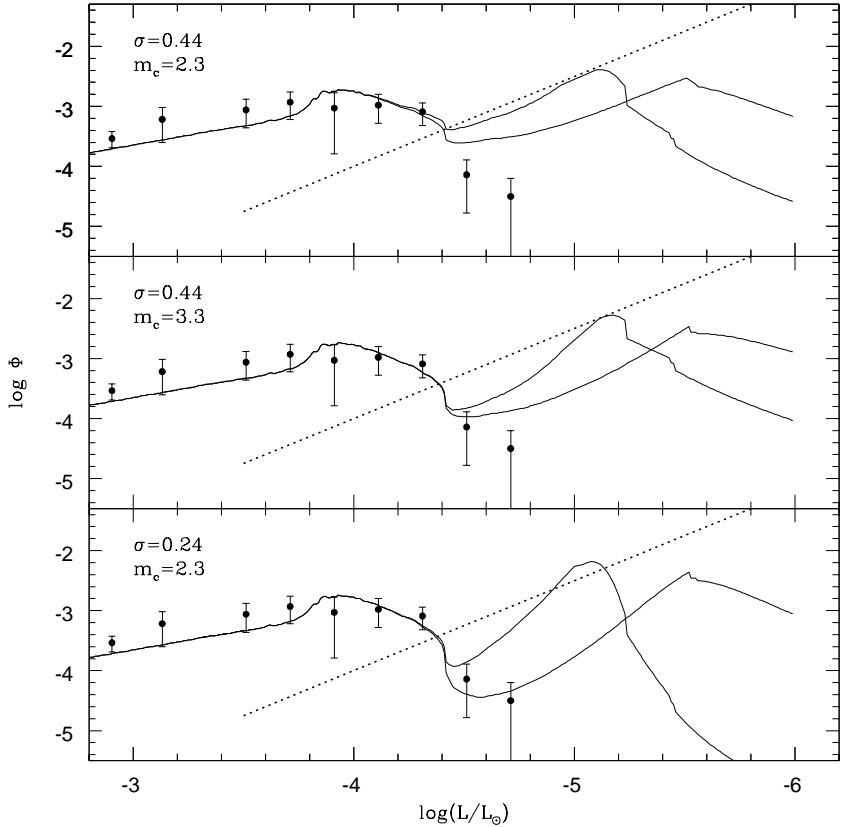,height=12.0cm}
\end{center}
\caption[]{The WDLF for the disk + halo.  The data points are the observed 
disk WDLF from Liebert et al. (\cite{liebetal88}); the dotted line represents
an upper limit based on the lack of white dwarfs on the HDF (assuming a
magnitude limit of $V=28$.  Solid lines represent theoretical WDLFs for DA
white dwarfs with a disk age of 9 Gyr, and halo ages of 12 Gyr and 14 Gyr.
Older models reach lower luminosities.  The three panels represent different
parameters for the IMF.  In comparison with the top panel, the middle panel
shows results for an IMF with a higher central mass, while the bottom panel
represnts an IMF with a much narrower width.  All models assume a halo white
dwarf mass equal to 25\% of the total mass of the dark halo.}
\end{figure}

The top panel of Figure 8 shows the model WDLF using the fiducial parameters
for the IMF.  While the WDLF falls below the HDF limit for both ages, the
lowest luminosity disk WDLF points are clearly inconsistent with the theory.
Shifting the IMF to higher masses, as illustrated by the middle panel of 
Figure 8, helps drop the theoretical WDLF near the disk cutoff, but is still
inconsistent with the observations.  The bottom panel illustrates that the
only way for a halo WDLF to satisfy the constraints of the disk observations
and the HDF is to narrow the IMF.  Bringing the width parameter down to
$\sigma=0.24$ satisfies both constraints for a halo age of 14 Gyr and older.
As the trend in Figure 8 shows, a younger halo with a narrower distribution 
might satisfy the disk constraints, but would then rise above the HDF upper
limit at slightly lower luminosities than the disk cutoff.  Another point to
consider is that the models used to construct the WDLFs in Figure 8 do not
include the effects of fractionation of the crystallizing white dwarf cores.
Graff et al. (\cite{grafetal98}) point out that the such an effect increases
the age constraints by about 2 Gyr.

There are a variety of other ways to compute the theoretical halo WDLF to
address the observations.  Again, the reader is urged to consult the papers
by Tamanaha et al. (\cite{tamaetal90}), Adams \& Laughlin (\cite{adalau96})
and Graff et al (\cite{grafetal98}) for more details of the influence of
various parameters on the halo white dwarf luminosity function.

The results presented in Figure 8 are intended to be representative and
schematic.  They show the essential results of the lack of finding white
dwarfs on the HDF.  First, the lack of white dwarfs on the HDF is not a
surprise given what we have learned from studies of the white dwarf
luminosity function of the galactic disk.  Second, the contribution of
white dwarfs to the Milky Way's massive dark halo can be limited based on this
lack of detection.  For reasonable parameters for the halo white dwarf
population, it is difficult for this null result on the HDF (and the disk
WDLF) to be consistent with a halo comprised of more than 25\% white dwarfs
by mass.  Third, if the limits of the HDF (or other very deep surveys) can 
be pushed fainter, the MACHO results require that some white dwarfs 
should be found (probably with an $M_{\rm bol}$ of about 17.5 to 19).
Finally, we can hope to learn a great deal about halo white dwarfs by
their discovery in abundance, even if it is not a chore for which HST is
ideally suited. 

\subsection{Postscript: white dwarfs on the HDF -- sort of}

In closing, the results of the WDLF studies of the disk of our galaxy, along
with the newer work on white dwarfs in the halo, allow us to estimate the
total number of white dwarfs in the Milky Way.  Assuming a uniform
distribution of white dwarf stars in the galactic disk, the local density
implies that there are approximately $3\times 10^9$ white dwarfs in the disk.
We can then use the luminosity function of the disk to estimate the total
luminosity from all of these white dwarf stars as approximately $3\times 10^6
L_{\odot}$. Thus, to ballpark accuracy, white dwarfs contribute a fraction of
about $3\times 10^{-5}$ of the total photon luminosity of our galaxy.

If we take the Milky Way as representative of the galaxies seen on the HDF,
we can ask how many photons that were counted by the HDF originated on a
white dwarf in a galaxy.  Given the number of galaxies on the HDF and the
total photon count, approximately 1 photon per galaxy came from a white
dwarf.  Therefore, white dwarfs have indeed been detected by the HDF; the
trick remains identifying which of those photons came from which white dwarf!

\begin{acknowledgments}
It is a pleasure to thank the organizers of this workshop for their help at
all stages of this review.  Matt Wood graciously provided many white dwarf 
evolutionary tracks for use in illustrating the properties of the white dwarf
luminosity function, and Russ Lavery assisted in the estimate of the white
dwarf photon flux on the HDF.  Partial support for this work also came 
through the NASA Astrophysics Theory Program (Grant NRA-96-04-GSFC-052) and
from an NSF Young Investigator award  (Grant AST-9257049 to Iowa State 
University).
\end{acknowledgments}


\end{document}